\documentclass[prb,twocolumn]{revtex4}

\usepackage{amsfonts}
\usepackage{amsmath}
\usepackage{amssymb}
\usepackage{graphicx}

\begin{document}

\title{In-plane hole density in (Ca$_{0.1}$La$_{0.9}$)(Ba$_{1.65}$La$_{0.35}$%
)Cu$_{3}$O$_{y}$ ; nuclear resonance study over the full doping range}
\author{Amit Kanigel}
\altaffiliation[Current Address: ]{Department of Physics, University of Illinois at Chicago, IL 60607}
\author{ Amit Keren}
\affiliation{Technion - Israel Institue of Technology.}

\begin{abstract}
We report in-plane $^{63}$Cu Nuclear Magnetic Resonance measurements for a
series of fully enriched (Ca$_{0.1}$La$_{0.9}$)(Ba$_{1.65}$La$_{0.35}$)Cu$%
_{3}$O$_{y}$ powder samples, which belong to the YBCO family, but the doping
could vary from very underdoped to extremely overdoped. From these
measurements we determine the average nuclear quadrupole resonance frequency 
$\nu _{Q}$, and its second moment $\Delta \nu _{Q}$, both set by the
in-plane hole density $n$, as a function of oxygen level $y$. We find that
in the overdoped side $n$ is saturated, but $\Delta \nu _{Q}$ rapidly
increases with increasing $y$. The relevance of these results to the
increasing penetration depth in overdoped cuprates is discussed.
\end{abstract}

\date{\today }
\maketitle

A common feature in the phase diagram of all cuprate high temperature
superconductors (HTSC) is the dome-shaped curve of the $T_{c}$ versus
doping, divided by an optimal doping point, and different properties on each
side. For example: the underdoped samples have a linear
temperature-dependent resistivity, while the overdoped samples seem to act
like normal metals with $T^{2}$-dependent resistivity. A pseudo-gap is found
in the underdoped side, but is absent from the overdoped side. The magnetic
penetration depth $\lambda $ has a minimum at optimal doping \cite%
{UemuraNat, Niedermayer,Locquet,oursolidstate}. In contrast Hall
measurements \cite{ando_1} done in single crystals of La$_{2-x}$Sr$_{x}$CuO$%
_{4}$ (La214), for example, show a smooth increase in the hole density with
increasing doping. This peculiar behavior led to the general belief that the
underdoped and overdoped regimes should be treated by different theories.
While the overdoped side is believed to adhere to mean field theory similar
to the BCS, the presence of strong electronic correlation in the underdoped
side motivated the development of exotic theories for this regime. Yet, few
authors found in the past that contrary to common belief, the hole density
is saturated in the overdoped side; for example in La214 \cite{UchidaPRB91},
Bi$_{2}$Sr$_{2}$CaCu$_{2}$O$_{8+\delta }$ (Bi2212) and Tl$_{2}$Ba$_{2}$CuO$%
_{6+\delta }$ (Tl2201) \cite{Puchkov}, and HaBa$_{2}$CuO$_{4+\delta }$
(Ha124) \cite{Puzniak}. This possibility could change the way we think about
overdoped cuprates. 
However, up until now, it could not be tested for the very popular and most homogenous 
YBa$_{2}$Cu$_{3}$O$_{y}$ (Y$123$) system since
it could not be overdoped without introducing disorder, e.g., by doping with Ca.

In this letter we overcome this problem by investigating the fully enriched
cuprate system (Ca$_{0.1}$La$_{0.9}$)(Ba$_{1.65}$La$_{0.35}$)Cu$_{3}$O$_{y}$
(CLBLCO). This system is unique in the sense that doping can vary across the
full range, from the very underdoped to the extreme overdoped, without any
structural changes \cite{clblco}. The system belongs to the Y$123$ family,
but it is tetragonal over the entire doping range, with no preferred
direction for the CuO chains and no chain ordering. Doping is controlled
only by the oxygen level \cite{clblco}. We determine the doping level by
extracting the nuclear quadrupole resonance parameter $\nu _{Q}$ from $^{63}$%
Cu Nuclear Magnetic Resonance (NMR) measurements. The Cu, with its spin $3/2$
nuclei, is directly coupled to charge degrees of freedom via the electric
field gradient (EFG), and $\nu _{Q}$ is a measure of this coupling. $\nu
_{Q} $, in turn, depends linearly on the hole density \cite{Asayama} $n$
according to 
\begin{equation}
\nu _{Q}=An+{}\nu _{Q}^{0}  \label{nuqvsn}
\end{equation}%
where $A$ and $\nu _{Q}^{0}$\ are doping-independent. This linear dependence
was demonstrated for various compounds such as Y123 \cite{Yasuoka}, La$214$ 
\cite{Zheng}, and Ha124 \cite{Gippius}. Therefore, Eq. \ref{nuqvsn}\ and the
ability of NMR to detect the in-plane cooper [Cu(2)] $\nu _{Q}$ selectively,
will allow us to determine the carrier concentration in the overdoped regime.

\begin{figure}[t]
\begin{center}
\includegraphics[width=9cm]{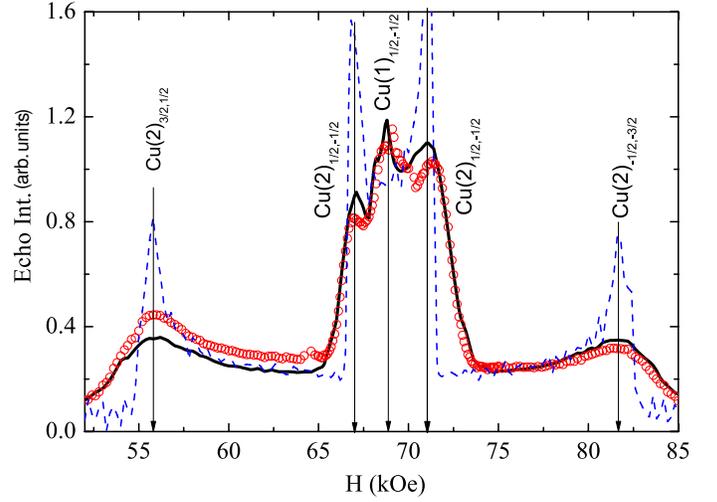}
\end{center}
\caption{ (Color online) Symbols: The spectra of the optimally doped CLBLCO sample ($%
y=7.135$). Dashed line:  Powder line shape reconstructed using the parameters
found by solving Eqs.~\protect\ref{Peaks}. Solid line:  Best fit to the data by
taking into account a distribution of $\protect\nu _{Q}$.}
\label{FullScane}
\end{figure}

The measurements were done on powder samples since CLBLCO is not orientable.
Their preparation is described in Ref.~\cite{clblco}. The oxygen content was
measured by double Iodometric titration. The accuracy of this method in
enriched CLBLCO is about 0.01. We measured seven different samples in the
normal state at 100~K. The most overdoped sample is not superconducting. The
NMR measurements were done by sweeping the field in a constant applied
frequency $f_{\text{app}}=77.95$ MHz using a $\pi /2$ - $\pi $ echo
sequence. The echo signal was averaged $100,000$ times and its area
evaluated as a function of field. The full spectrum of the optimally doped
sample ($y=7.135$) is shown in Fig.~\ref{FullScane}. Four peaks associated
with the plane Cu(2) and one peak from the Cu(1) can be clearly seen.

\begin{figure}[t]
\begin{center}
\includegraphics[width=9cm]{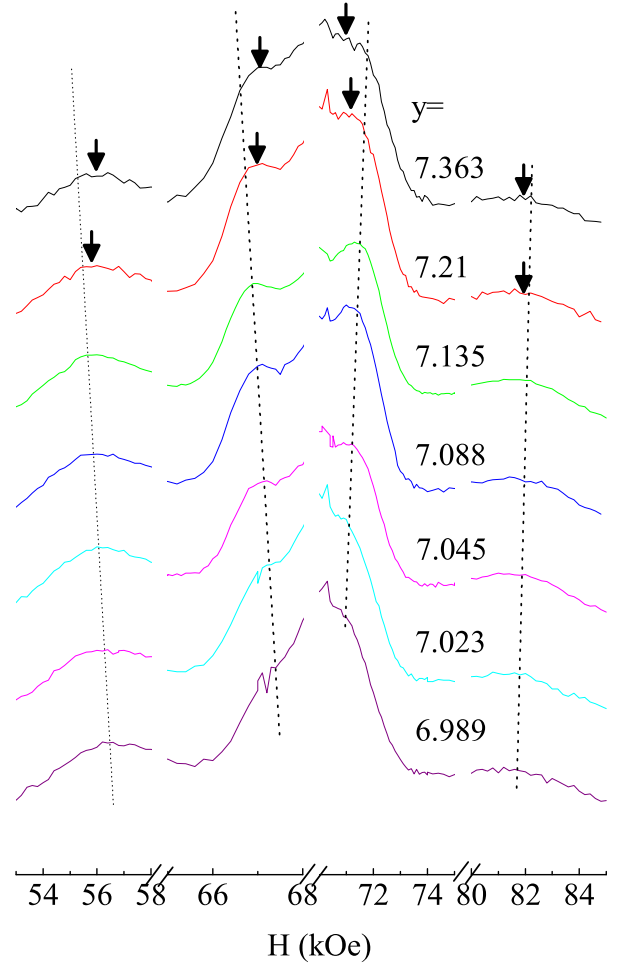}
\end{center}
\caption{(Color online) Spectra for the 7 CLBLCO samples. Position of the four peaks
associated with the plane Cu as function of doping from under to optimal ($%
y=7.135$) doping are shown by the dotted lines. These lines are extended
further to show the opposite peak position evolution for overdoped samples.
Their peak position are shown by arrows.}
\label{AllSamples}
\end{figure}

A zoom on the main features of the Cu(2) signal of all seven samples is
depicted in Fig.~\ref{AllSamples} (note the three axis breakers). The
evolution of the main peaks with doping from underdoped to the optimally
doped ($y=7.135$) samples is highlighted by the dotted line. This line is
extended to the overdoped samples to demonstrate their opposite peak
position evolution with doping. For overdoped samples, arrows indicate the
peak positions. The opposite peak position evolution of the overdoped
samples is the main analysis independent finding of this work. It shows that
charge density in the CuO$_{2}$ plane is not increasing in the overdoped
side.

The Cu spin Hamiltonian can be written as \cite{Slichter}: 
\begin{equation}
\mathcal{H}/h=-\nu _{l}\mathbf{I\cdot (1+K)\cdot \hat{H}}+\frac{\nu _{Q}}{6}%
[3\mathbf{I}_{z}^{2}-\mathbf{I}^{2}+\eta (\mathbf{I}_{x}^{2}-\mathbf{I}%
_{y}^{2})]  \label{H}
\end{equation}%
where $\nu _{l}=(^{63}\gamma /2\pi )H$, $\hat{\mathbf{H}}$ is a unit vector
in the direction of the field, $\mathbf{K}$ is the shift tensor, and $\eta $
is the asymmetry parameter of the EFG. In the absence of magnetic field,
there is only one transition frequency given by $f=\nu _{Q}\sqrt{1+\eta /3}$%
, so $\nu _{Q}$ cannot be separated from $\eta $, and the use of the
magnetic field is essential. This field, applied in the direction $\theta $
and $\phi $ with respect to the principal axis of the EFG, lifts this
degeneracy and three transition frequencies $\nu _{m}(H,\theta ,\varphi )$
are expected: a center line which correspond to the $1/2\rightarrow -1/2$
transition ($m=0$), and two satellites which corresponds to the $%
3/2\rightarrow 1/2$ ($m=1$) and $-1/2\rightarrow -3/2$ ($m=-1$) transitions.
Expressions for $\nu _{m}(H,\theta ,\varphi )$ up to second order
perturbation theory in $\nu _{Q}$ for completely asymmetric EFG and shift
tensors are given in Ref.~\cite{Taylor}.

For calculating a field-swept powder spectrum $P(H)$ one has to evaluate%
\begin{equation}
P_{m}(H)\propto \sum_{m}\int M^{2}\delta \lbrack f_{\text{app}}-\nu
_{m}(H,\theta ,\varphi )]d\Omega  \label{trans_cond}
\end{equation}%
where $M$ is a matrix element. This spectrum is known to have four peaks
(two emerge from $m=0$) at fields given \cite{Taylor,KerenPRB98} by 
\begin{subequations}
\label{Peaks}
\begin{align}
f_{\text{app}}& =\frac{1}{2}\nu _{Q}(1-\eta )+\nu _{ls} \\
f_{\text{app}}& =\frac{\nu _{Q}^{2}}{48\nu _{lc}}(9-6\eta +\eta ^{2})+\nu
_{lc} \\
f_{\text{app}}& =-\frac{\nu _{Q}^{2}}{3\nu _{hc}}(1-\eta )-\frac{%
4(K_{\bot}-K_{z})^{2}\nu _{hc}^{3}}{3\nu _{Q}^{2}(\eta -3)^{2}} \\
& -\frac{2K_{\bot}(\eta -2)+K_{z}(\eta -5)+3(\eta -3)}{3(3-\eta )}\nu _{hc} 
\notag \\
f_{\text{app}}& =-\frac{1}{2}\nu _{Q}(1-\eta )+\nu _{hs}
\end{align}%
where $\nu _{ls}$, $\nu _{lc}$, $\nu _{hc}$, $\nu _{hs}$ are the low
satellite, low center, high center, and high satellite peaks, respectively,
and we assume $K_{x}=K_{y}=K_{\bot }$ since CLBLCO is tetragonal. Usually
one can extract all the averaged Hamiltonian parameters from the position of
the peaks in the spectra by solving Eqs.~\ref{Peaks} numerically \cite%
{KerenPRB98}. However, we are interested in both the parameters and their
distribution. Therefore, we use a grid in the $(\theta ,\phi )$ space,
calculate the frequency $\nu _{m}(H,\theta ,\varphi )$ for every $m$, field,
and point on the grid, and add 1 to a histogram of $H$ when this frequency
equals $f_{\text{app}}$. The matrix elements are taken as unity. The
numerical summation approximates $P(H)$ in Eq.~\ref{trans_cond}.

This numerical approach is demonstrated in Fig.~\ref{FullScane} where the
four peaks associated with Cu(2) have almost axial symmetry ($\eta \sim $
0), and the fifth (middle) peak is from Cu(1) which has a completely
asymmetric EFG ($\eta \sim 1$). The dashed line in Fig.~\ref{FullScane} is
the powder line shape calculated after the parameters for Cu(2) were
extracted by solving Eqs.~\ref{Peaks} for the plane site. The singularities
of the theoretical line agree with the peaks in the NMR data, as expected.
However, the overall shape of the calculated line does not resemble the data
at all, the reason being the distribution of the Hamiltonian parameters. The
main contribution to the width, of the order of a few MHz, is from a
distribution in $\nu _{Q}$, since the quadrupole interaction is the only
interaction of such magnitude in the system. To account for this
distribution we simulated the line shape by adding to the numerical
evaluation of Eq.~\ref{trans_cond} a loop over $200$ values of $\nu _{Q}$
drawn from a normal distribution with a width $\Delta \nu _{Q}$. We also
added the contribution of the chain site, with $\eta \sim 1$ and removed the 
$K_{x}=K_{y}=K_{\bot }$ constrain, to take into account possible local
orthorhombic distortions. We searched for the best fit to the data by $\chi
^{2}$ minimization using a simplex code \cite{NumericalRec}. The result for
the optimally doped sample is shown as the solid line in Fig.~\ref{FullScane}.

In Fig.~\ref{Positions} we show the position of the four peaks (shown in
Fig.~\ref{AllSamples}) as a function of doping. The separation between the
satellites increases with doping up to the optimal doping point and
decreases as the system is overdoped. The evolution of the width of the
lines can be seen by examining the peaks of the center lines in Fig \ref%
{AllSamples}.

By solving Eqs.~\ref{Peaks}, we calculate $\nu _{Q}$, $\eta $, $K_{z}$ and $%
K_{\bot }$ for all the samples. The calculated $\nu _{Q}$s are
shown in Fig. \ref{Nuq}. All parameters are used as the initial guess in the
fitting program. The results for fitted $\nu _{Q}$ and $\Delta \nu _{Q}$ are
also shown in Fig.~\ref{Nuq}. The solid lines in this figure are guides for
the eye. The width of the lines does not allow us to determine $\eta $, $%
K_{z}$, $K_{x}$ and $K_{y}$ very accurately. In general the Cu(2) $\eta $ is
always very small for all the samples, ranging from 0.015 for the optimal
doped sample and up to 0.07 for the underdoped sample, in agreement with
Y123 \cite{eta}. We have also attempted to fit the data starting from
arbitrary initial guess. This gave somewhat different fit values. The error
bars in Fig.~\ref{Nuq} are estimated from the different fit procedures.
These error bars are larger by a factor of 4 than the error bars calculated
by the fitting program itself.

\begin{figure}[t]
\begin{center}
\includegraphics[width=9cm]{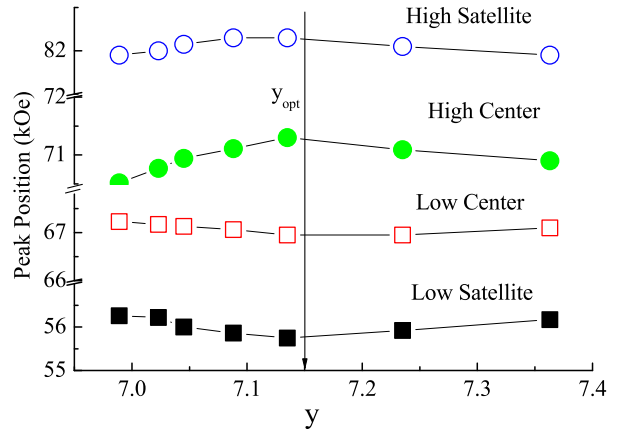}
\end{center}
\caption{(Color online) Position of the 4 peaks associated with the plane Cu as function of
doping. }
\label{Positions}
\end{figure}

The location of the simulated peaks agrees with the data for all samples,
but the simulation program yields a higher $\nu _{Q}$ than extracted from
Eqs.~\ref{Peaks}. The reason is that the theoretical peaks for zero $\Delta
\nu _{Q}$ are not symmetric (see dashed line in Fig.~\ref{FullScane}). By
introducing a normal distribution of $\nu _{Q}$ the peaks tend to shift
towards the center of the line resulting in what seems to be a smaller
average $\nu _{Q}$. The computer fit program corrects for this effect by
selecting a higher average $\nu _{Q}$ than the initial guess (based on Eqs.~%
\ref{Peaks}). We believe that the fit results represent the data better than
the calculation that is based on the peaks alone.

The results indicate that $\nu _{Q}$ grows linearly with doping in the
underdoped side of the phase diagram in agreement with other compounds \cite%
{Yasuoka,Zheng,Gippius}. In the overdoped side $\nu _{Q}$ grows
very slowly with doping and saturates; it does not decrease as one might
think looking at the peaks position only. This is due to the increase in $%
\Delta \nu _{Q}$. The width decreases with doping, having a minimum at
optimal doping, and increases quite sharply in the overdoped side. We
emphasize, that the main result, a deviation from the linear dependence of $%
\nu _{Q}$ in the overdoped side, is observable in the raw data itself. The
same result is supported both by the exact solution of Eqs.~\ref{Peaks} and
by the numerical fits.

The fact that there is a sudden change in the $\nu _{Q}$ versus $y$ relation
suggests a change in the doping mechanism itself. In CLBLCO, like in Y123,
the doping is done by adding oxygen ions into the plane of chains, which are
initially only half full. Our result suggests that in the underdoped region
a constant part of the introduced holes goes into the CuO$_{2}$ plane.
 After crossing the optimal doping point, most
of the holes remain in the chains. 

Recent density functional calculations
 of doping and its effect on the EFG \cite{StollUnKnown} in YBCO show that
both $\nu _{Q}$ and the doping should go through a maximum as a function of
the oxygen level. It should be noted that resistivity and Hall measurements
in YBCO will be less sensitive to the change in the plane doping mechanism
since above optimal doping the chain layers start to conduct.

\begin{figure}[tbp]
\begin{center}
\includegraphics[width=9cm]{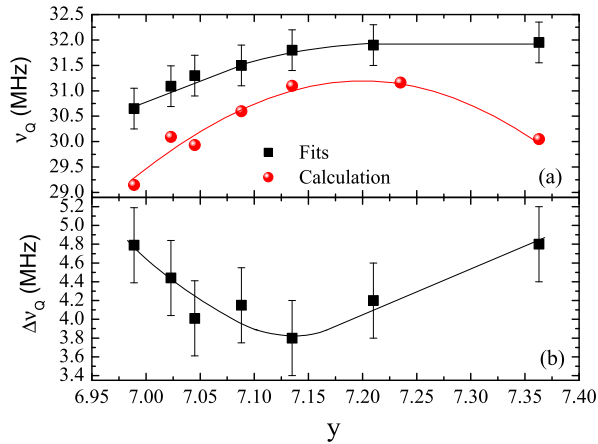}
\end{center}
\caption{ (Color online) (a) $\protect\nu _{Q}$, the mean value of the quadruple frequency,
versus doping as obtained by fit and calculation described in the text. (b) $%
\Delta \protect\nu _{Q}$, the width of the normal distribution around $%
\protect\nu _{Q}$ obtained by the fit.}
\label{Nuq}
\end{figure}

Thus, above optimal doping the energetic cost of adding a free hole to the
CuO$_{2}$ plane increases. Trying to overdope CLBLCO results in transfer of
charge into the chains. Moreover, attempts to overdope the sample, leads to
higher degree of disorder in the plane, manifested in the increase of $%
\Delta \nu _{Q}$. This is the main conclusion of this work.

Finally, it is interesting to compare the NQR results with previous magnetic
penetration depth, $\lambda $, measurements \cite{oursolidstate}. $\lambda $ was
determined by transverse field muon spin rotation where the relaxation rate $%
\sigma $ is proportional to $\lambda ^{-2}$ [ref \cite{MuSRBook}]. In CLBLCO $%
\sigma $ is proportional to $T_{c}$ in both the under and over doped sides,
with the same proportionality constant of all HTSC, obeying the Uemura
relation \cite{oursolidstate}. Within the clean limit BCS theory framework, at $T=0$
all the normal state carriers should pair and condensate giving $\sigma
\propto n_{s}/m^{\ast }$ where $n_{s}=n/2$ and the effective mass is just
twice the effective mass of the quasi-particle in the normal state.
Therefore, the saturation of the number of holes in the plane above optimal
doping cannot explain the reduction in $\sigma $ alone. Clearly the
explanation must involve a combination of the saturation of the normal state
carrier density and some other mechanism. Few
suggestion for the nature of this mechanism were proposed, for example,
phase separation and creation of metallic droplets \cite{UemuraNat}, and pair
breaking \cite{Niedermayer}. The fact that we find an increase in the width
of the charge distribution in the planes might be an important ingredient in
the correct theory explaining the peculiar behavior of the superfluid
density in the overdoped side.

In summary, we have managed to measure the average quadrupole resonance
parameter $\nu _{Q}$ and the width of its distribution $\Delta \nu _{Q}$,
for the in-plane Cu, in a system belonging to the YBCO family but where doping can
vary from heavily underdoped to extreme overdoped. We find that $\nu _{Q}$
increases with oxygen doping in the underdoped side but is saturated in the overdoped side. In
contrast, $\Delta \nu _{Q}$ has a minimum at optimal doping.

We acknowledge useful discussion with Y. Eckstein and the help of Arkady Knizhnik with the Iodometric titation. This work was funded by
the Israeli Science Foundation.

\end{subequations}


\begin{thebibliography}{99}
\bibitem{UemuraNat} Y.J. Uemura, A. Keren, L.P. Le, G.M. Luke, W.D. Wu, Y. Kubo, T.Manako, Y. Shimakawa, M. Subramanian, J. L. Cobb  and  J. T. Markert \newblock{Nature}, \textbf{%
364} 605 (1993).

\bibitem{Niedermayer} Ch. Niedermayer, C. Bernhard, U. Binninger, and H. GlŸckler
\newblock{Phys. Rev.
Lett.}, \textbf{71} 1764 (1993).

\bibitem{Locquet} J.P. Locquet, Y. Jaccard, A. Cretton, E. J. Williams, F. Arrouy, E. MŠchler, T. Schneider,    O. Fischer and P. Martinoli \newblock{Phys. Rev B}, 
\textbf{54}, 7481 (1996).

\bibitem{oursolidstate} A. Keren, A. Kanigel, J. S. Lord and A. Amato 
\newblock{Solid State
Comm.}, \textbf{126} 39 (2003).

\bibitem{ando_1} Yoichi Ando, A. N. Lavrov, Seiki Komiya, Kouji Segawa, and X. F. Sun \newblock{Phys. Rev. Lett.}, 
\textbf{87}, 017001 (2001).

\bibitem{UchidaPRB91} S. Uchida, T. Ido, H. Takagi, T. Arima, Y. Tokura, and S. Tajima \newblock{Phys. Rev. B}, 
\textbf{43} 7942 (1991).

\bibitem{Puchkov} A. V. Puchkov, P. Fournier, T. Timusk, and N. N. Kolesnikov \newblock{Phys. Rev. Lett}, 
\textbf{77}, 1853 (1996).

\bibitem{Puzniak} R. Puzniak, R. Usami and H. Yamauchi \newblock{Phys. Rev. B} \textbf{\
53}, 86 (1996).

\bibitem{clblco} D. Goldschmidt, A. Knizhnik, Y. Direktovitch, G. M. Reisner, and Y. Eckstein \newblock{Phys Rev. B} \textbf{\ 49},
15928 (1994).

\bibitem{Uemurasolidstate} Y.J. Uemura \newblock{Solid State Comm.} \textbf{120} , 347
(2001).

\bibitem{Asayama} Kunisuke Asayama, Yoshio Kitaoka, Zheng Guo-qing and Kenji Ishida
\newblock{ Prog. Nuc.
Mag. Reso. Spec.}, \textbf{28}, 221 (1996).

\bibitem{Yasuoka} H.Yasuoka, in Spectroscopy of Mott Insulator and
Correlated Metals, edited by A. Fujimori and Y. Tokura, Solid State
Sciences, Vol. 119 (Springer-Verlag, Berlin, 1995), p. 213.

\bibitem{Zheng} Guo-qing Zheng, Yoshio Kitaoka, Kenji Ishida and Kunisuke Asayama, J. Phys. Soc. Jpn. \textbf{64},
2524 (1995).

\bibitem{Gippius} A. A. Gippius, E.V. Antipov, W. Hoffmann and K. Luders, Physica C \textbf{276}, 57
(1997).

\bibitem{Slichter} C.P. Slichter \newblock{Principles of Magnetic Resonance}%
, Harper and Row, New York, (1963).

\bibitem{Taylor} J. F. Baugher, P.C. Taylor, T. Oja and P.J. Bray 
\newblock{
J. of Chem. Phys.}, \textbf{50}, 4914 (1969).

\bibitem{KerenPRB98} A. Keren, P. Mendels, M. Horvati?, F. Ferrer, Y. J. Uemura, M. Mekata and T. Asano \newblock{Phys. Rev. B}  \textbf{57},
10745 (1998).

\bibitem{NumericalRec} W.H. Press, B.P. Flannery, S. Teukolsky and W.T. Vetterling,  
\newblock{numerical
Recipes in C}, Cambride University Press (1988).

\bibitem{eta} C. H. Pennington, D. J. Durand, D. B. Zax, C. P. Slichter, J.
P. Rice, and D. M. Ginsberg, Phys. Rev. B \textbf{37}, R7944 (1988); T.
Shimizu \textit{et al.}, Journal of the Physical Society of Japan, \textbf{57%
}, 2494 (1988).



\bibitem{StollUnKnown} E. P. Stoll and P. F. Meier, Private Communication.



\bibitem{MuSRBook} Muon Science: Muons in Physics, Chemistry and Materials,
Eds S. L. Lee, S. H. Kilcoyne and R. Cywinski (Institute of Physics,
London), 1999.
\end{thebibliography}
\end{document}